\documentclass[aps,prl,twocolumn,showpacs,superscriptaddress,groupedaddress,floatfix]{revtex4-1}  
\usepackage{graphicx,epsfig}
\usepackage{float}
\usepackage{epstopdf}
\usepackage{hyperref}
\usepackage{mdwlist}
\usepackage{amssymb}
\usepackage{amsmath}
\usepackage{amsfonts}
\usepackage{textcomp}
\usepackage{amssymb}
\usepackage{placeins}
\usepackage{textcomp}
\usepackage[utf8]{inputenc}
\usepackage[caption=false]{subfig} 

\begin{document}

\title{A Langevin model for complex cardiological time series}

\author{C.E.C. Galhardo$^1$, B. C. Coutinho$^2$, T.J.P.Penna$^3$, M.A. de Menezes$^3$  and P.P.S. Soares$^4$ } 
  
  \affiliation{$^1$ Materials Metrology Division, National Institute of Metrology, Quality and Technology, Duque de Caxias, RJ ,Brazil}
  \affiliation{$^2$ Petrobrás, Cenpes, Rio de Janeiro, RJ, Brazil}
  \affiliation{$^3$ Instituto de Física,  Universidade Federal Fluminense,  Niteroi, RJ ,Brazil } 
  \affiliation{$^4$ Instituto Biomédico,  Universidade Federal Fluminense,  Niteroi, RJ, Brazil }

\begin{abstract}

There has been considerable efforts to understand the underlying complex dynamics in physiological time series.  
Methods originated from statistical physics revealed a non-Gaussian statistics and long range correlations in those signals. 
This suggests that the regulatory system operates out of equilibrium. Herein the complex fluctuations in blood pressure 
time series were successful described by physiological motivated Langevin equation under a sigmoid restoring force with 
multiplicative noise. 

\end{abstract}


\maketitle
\date{\today}

\textit{Introduction -} The autonomic nervous system is able to maintain life signals at safe levels by 
the action of a pair of nerve branches, 
called sympathetic and parasympathetic. While the sympathetic prepares our body for ''flight or fight'' 
the parasympathetic (or vagal) 
is considered as a ''rest and digest'' system. 
In many cases, to achieve the homeostatic optimal levels, these systems have a competitive approach: while one start up 
an physiological reaction the other one suppress it \cite{AmJPharEdu:71:78}. 

Homeostasis depends on the blood flow according to the metabolic demands of each body part. The exchange of nutrients 
and metabolites occurs when blood flows through capillary channels. The perfusion either into or out the 
capillary depends on the blood pressure. Adequate levels of blood pressure are controlled by several mechanisms that can be 
classified according to the response delay: the long term and short term control.

In order to maintain homeostasis the body automatically responds to changes. These responses are called reflexes. 
The principal short term reflex regulation of arterial blood pressure is the baroreflex. Stretch-sensitive mechanoreceptors 
are located in the carotid sinus and aortic arch connected to
the brainstem, or nucleus tractus solitarii, by the glossopharyngeal and vagal nerves. After integrating the 
afferent signals the central nervous system, in turn, 
excite/inhibit the vagal branch if the pressure is high/low enough, closing the circuit for what 
can be regarded as a self-inhibitory feedback \cite{book_phys}.

In addition, blood pressure may vary to adapt different physiological conditions 
such as exercise  \cite{ExpPhysiol:91:79} or pregnancy \cite{AJP_RICP:258:R1417}  and in certain disease states such as hypertension \cite{Hyper:31:68}. 
The optimal level of blood pressure 
must be risk adjusted: cannot be high enough to cause structural damage and cannot be low enough to hinder the nutrient flow. 

It has been suggest that the underlying cardiac control system can be characterized as a complex system. 
Indeed, heart and blood pressure present fractal time series with long range correlations and 
non-Gaussian distributions \cite{PhysRevLett:70:1343} and 
undergoes a breakdown of critical characteristics like a continous second ordem phase transtion \cite{PhysRevLett.95.058101}.

In this letter, we propose an analytically tractable stochastic model 
for the baroreflex that capture the fractality and the non-Gaussian behavior in the blood pressure time series.
This model leads to solutions in terms of the so called \textit{q}-Gaussians, 
which is a well known function in the framework of Nonextensive Statistical Mechanics.
This theory have shown excellent results describing time series systems with fractal structure and 
an introduction and many examples of the theory can be found in \cite{tsallis2009introduction}.

\textit{Experiment - }  Two groups of adult male Wistar rats were analyzed: control rats 
and chronic sinoaortic denervated rats, animals surgically denervated $20$
days before measurements. Sinoaortic denervated was performed using the methods
described by Krieger et al \cite{CircRes:15:511}, and basically
consists of full disruption of the nerve fibers connecting the
the afferent signals from the baroreceptors to the brainstem, 
leading to hypertension, tachycardia, and an increase in blood pressure lability. 
Twenty days after the surgery, only the increase of blood pressure lability is usually observed \cite{JCardPhar:47:331}.
 This adaptation had been previously investigated in the light of 
detrended fluctuation analysis \cite{NJP:11:103005}. 

 Blood pressure was recorded from the left femoral artery for
$90$ minutes in conscious rats. Before the analog to digital conversion, blood pressure was low-pass
filtered (fc= $50$ Hz) for high-frequency noise removal, and recorded with a $2$kHz sampling
frequency.  Diastolic (minimum) values were detected after parabolic interpolation and 
signal artifacts were visually identified and removed.  A more detailed account of this 
experiment can be found in \cite{JCardPhar:47:331}. Since the measurements were done in awake conscious 
unrestrained rats some distortions in the blood pressure signal arise due to their movements.
To reduce this problem we discard series that show any kind of discontinuities. 

\begin{figure}[H]
\centering
 \begin{minipage}{0.99\columnwidth}
\centering

\includegraphics[width=0.9\columnwidth]{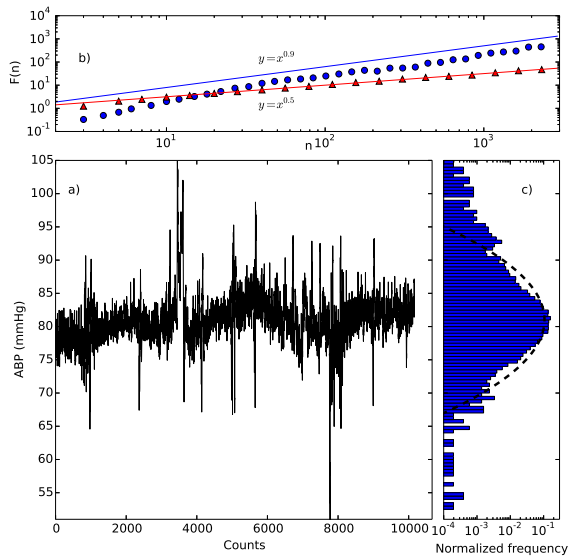}
\caption{Representative data of diastolic blood pressure from one animal. 
(a) a typical diastolic blood pressure record is presented.  (b)  
The histogram for the blood pressure values. A Gaussian distribution (black dash-line) with sample mean and sample variance 
was plotted with the histogram, highlighting the non-Gaussian behavior.
(c) DFA for diastolic blood pressure series. For long time scales the DFA exponent is $ \alpha = 0.93 $. 
We also applied DFA to shuffled data (red up triangles) 
obtaining $ \alpha \approx 0.5 $. The curves $\alpha=0.5 $ (red full line), $ 0.9 $ (blue full line) are plotted as guides 
to the eye.}
\label{fig:represent}
\end{minipage}
\end{figure}

\textit{Results and Discussion -} Figure \ref{fig:represent}a a typical diastolic blood pressure record from a control animal is presented. 
The histogram for the blood pressure values is presented in figure \ref{fig:represent}c.  A Gaussian distribution with sample mean and 
sample variance was plotted with the histogram, highlighting the non-Gaussian behavior. Furthermore, it clearly presents a positive skew. 
Despise the fact that the time series has a constant mean of $81.00 mmHg$ over a range of $10^4$ inter-beats, a complex pattern 
of fluctuations were observed. To characterize the stochastic dynamics of the blood pressure a detrended fluctuation analysis (DFA) was performed. 
The log-log plot of the fluctuation function $F(n)$ is presented in figure \ref{fig:represent}b. 
A crossover around $n=35$ separate two different behaviors. For long time scales the DFA exponent $\alpha = 0.93$ while a 
for short time scales $\alpha = 1.18$. This crossover had been discussed 
previously and it could be associated 
with autonomic nervous system control of arterial blood pressure \cite{NJP:11:103005,fuchs2010comparing}. To confirm the presence of long range correlations, 
the time series were randomly shuffled and DFA was applied. The shuffled time series present DFA exponent $\alpha\approx 0.5$,
as typical white noise, show at figure \ref{fig:represent}b in red, which demonstrates that long range correlations arises from the blood pressure 
control system.

External perturbations are continuously disrupting the cardiac system. 
In such noisy environment the autonomic nervous system 
must keep the blood pressure at acceptable levels by integrating chemical and mechanical input from afferents
to regulate the blood pressure. However, the neural transmission itself is noisy \cite{stein2005neuronal}. 
Information transmission along the axons is made by electrical signals. Those signals, called action potentials, are created by ion channels in 
a cell membrane. They travel down the axon to its end where the neurotransmitters are 
release in the synaptic gap. Those transmitters activate receptors in the post synaptic neuron \cite{white2000channel}. The noise 
source are diverse, for example, they could have physical background in thermodynamic or quantum mechanics, 
as it happens with sensor neurons, or could be built up in the cellular network \cite{faisal2008noise}. In this sense, the cardiac neural 
control system is under intrinsic and extrinsic noise. 

 \begin{figure}[H]
   \centering
 \begin{minipage}{0.99\columnwidth}
  \centering 
 
   \subfloat[Diastolic pressure distribution control rats]{
       \includegraphics[width=0.48\columnwidth,angle=0]{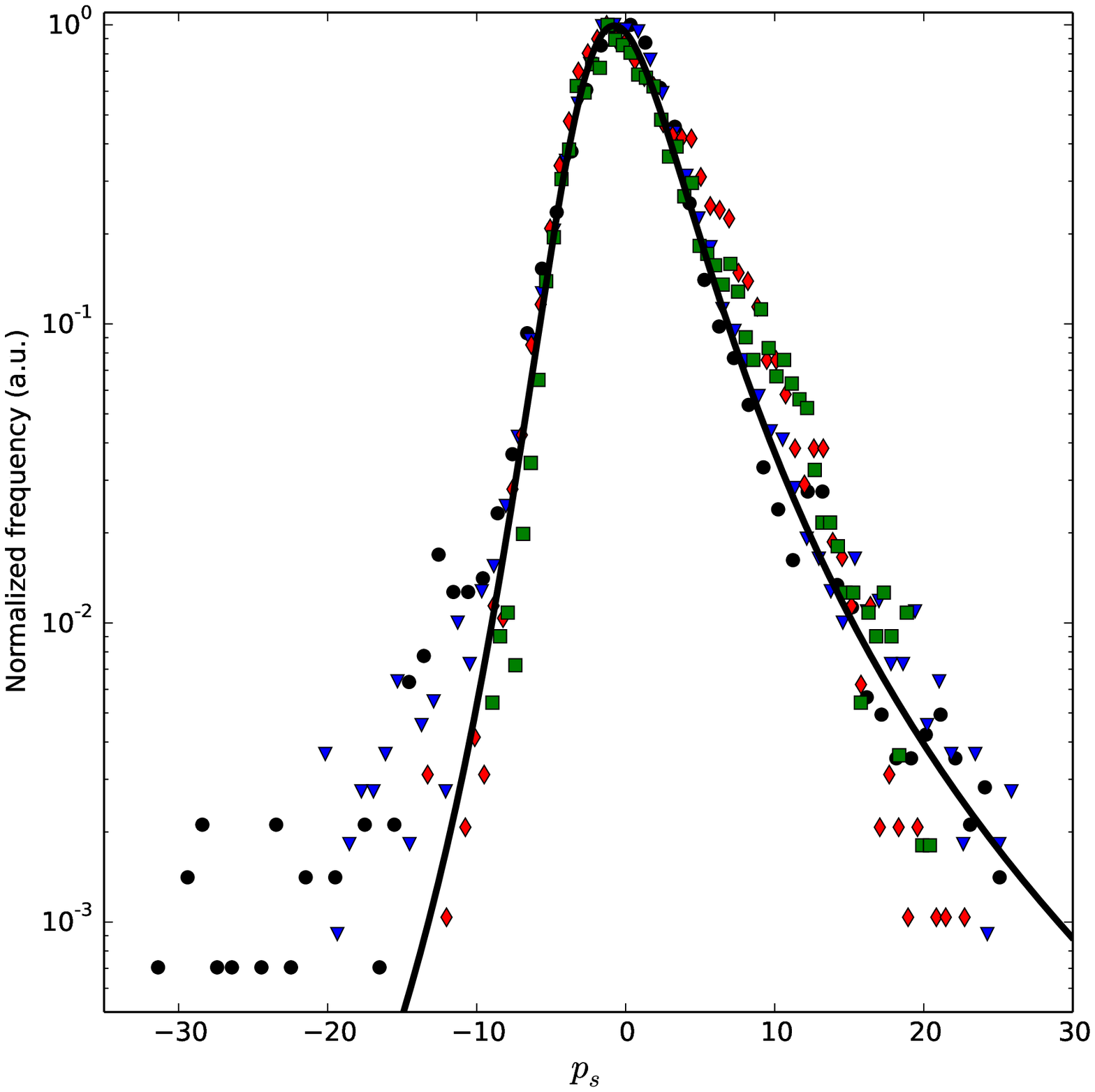}
       \label{fig:colapse_rats_ctr}
   }
   \subfloat[Diastolic pressure distribution denervated rats]{
       \includegraphics[width=0.48\columnwidth,angle=0]{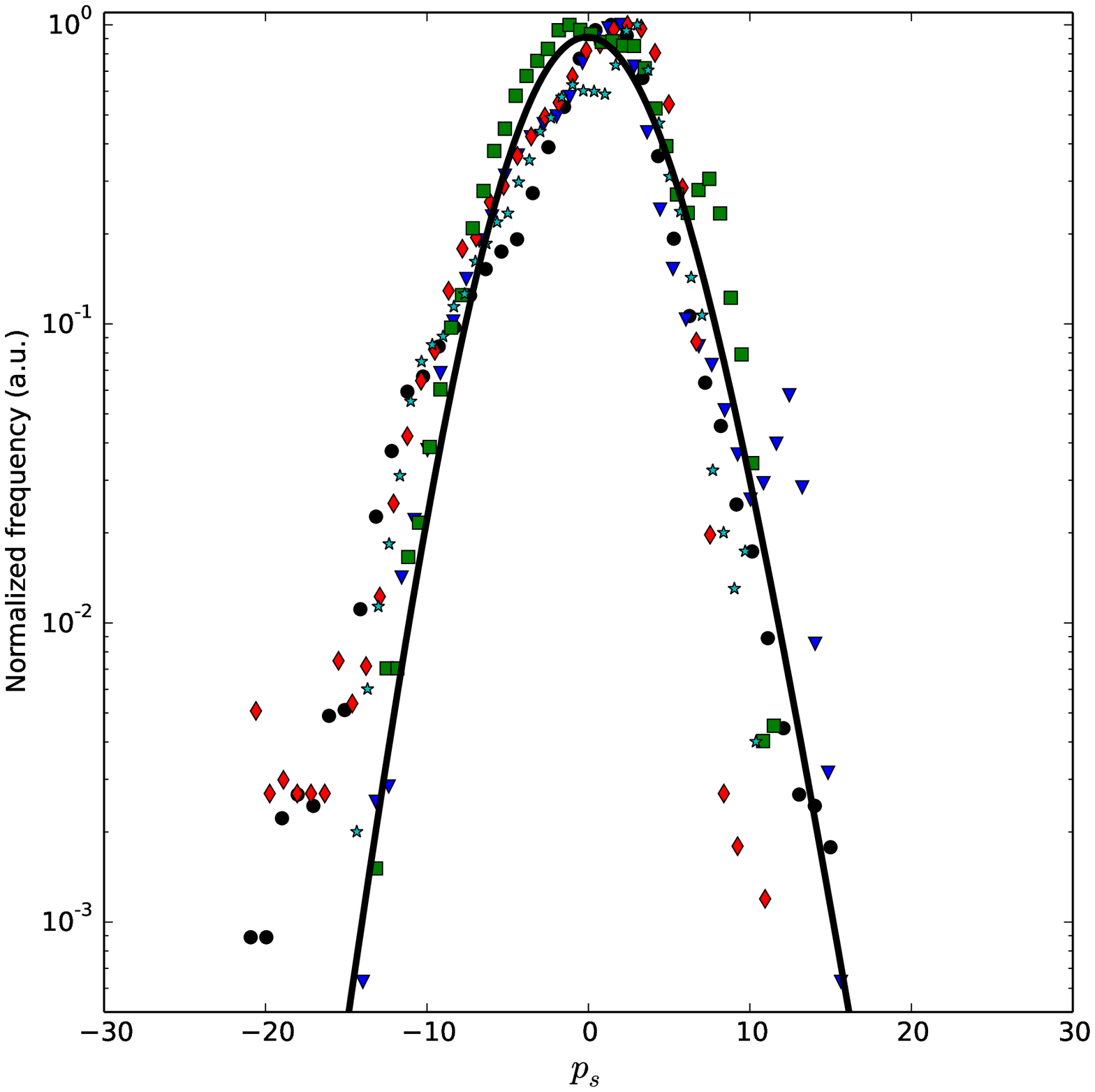}
       \label{fig:colapse_rats_20d}
   }

 \caption{Semi log histogram of rescaled experimental diastolic pressure data. Figure \ref{fig:colapse_rats_ctr} present the data of four control rats 
 while figure \ref{fig:colapse_rats_20d} shows the data of four denervated animals. All data where rescaled  $p_s = k_0(p - \tilde{p})/\sigma$, 
 where $k_0 = 4$, $p$ is the blood pressure data, $\tilde{p}$ is the median  and $\sigma$ is the sample standard 
 deviation. Despite the fact that the animals were unrestrained during the measurement, the rescaled data collapse. 
 In each figure the stable 
 solution of the model where plotted. For the control animals, figure \ref{fig:colapse_rats_ctr}, 
 a \textit{q}-Gaussian distribution 
 with $q=11/9$ were found. For the denervated rats,figure \ref{fig:colapse_rats_20d}, 
 a almost Gaussian distribution were found with $q=1.04$.}
 \label{fig:colapse}
 \end{minipage}
 \end{figure}

Synaptic signal transmission is found to be modeled as
a diffusive process 
wherein additive and multiplicative noise plays 
relevant role to describe the neuronal response \cite{boucsein2009dynamical,richardson2010firing}. Let $p = P - \tilde{p}$ 
where $P$ is the blood pressure and $\tilde{p}$ is a measure of central tendency, like the mean or the median. Then the neural control of blood pressure 
dynamics could be modeled as a Brownian particle under a restoring force of the baroreflex:

\begin{equation}
\frac{dp}{dt} = f(p) + g(p)\xi(t) + \eta(t) \label{eq:langevin_geral}.
\end{equation}

\noindent where $f(p)$ is the restoring force, $g(p)$  is the diffusion coefficient, $\xi(t)$ is the multiplicative noise 
and $\eta(t)$ is the additive noise. They both have zero mean and show Markovian correlations, 
 $\langle \xi_t \xi_{t'}\rangle = M \delta_{tt'} $ and $ \langle \eta_t \eta_{t'}\rangle = R \delta_{tt'}$. Physiologically $\eta(t)$ 
could be interpreted as external perturbations on cardiac system while $g(p)\xi(t)$ is a noise
arising from the neural transmission. 

Harris and Wolpert \cite{harris1998signal} propose an unifying optimal control theory for information process in motor systems. 
The theory is based on a single physiological assumption: the neural noise increases in variance with the size of the control signal. 
This assumption is then made by the model presented in this paper. Here the noise amplitude is 
signal dependent and it is proportional to the control system. Several complex systems, like financial and 
turbulence \cite{peinke1999uniform,ghashghaie1996turbulent}, has 
a linear drift coefficient and a quadratic state-dependent diffusion coefficient. In those system $f(p)= -\frac{dU(p)}{dp}$ and 
$g^2(p) \propto U(p)$. Analogously, the restoring force of the baroreflex could be expressed as above with
$U(p) = \frac{\tau}{2} g(p)^2$.Generally the baroreflex is represented by a sigmoid function \cite{kent1972mathematical}, 
$f(p)= A - \frac{C}{1+e^{-Bp}}$ where $A,B,C$ are the model parameters. Integrating the sigmoid one can found 
$g^2(p)= \frac{2}{\tau} \left[ -Ap + C \frac{\text{ln}(1+e^{Bp})}{B} \right] $
where $A < C$, $A>0$, and $B > 0$ must hold that $g^2(p)$ be strictly positive. 

Equation (\ref{eq:langevin_geral}) could be rewritten as Fokker-Planck equation using Ito calculus:

\begin{equation}
\frac{\partial F(p)}{\partial t} = - \frac{\partial [f(p)F(p)]}{\partial p} + 
\frac{\partial [G'(p)F(p)]}{\partial p} + G(p)\frac{\partial^2 [F(p)]}{\partial p^2} 
\label{eq:noise_induced}
\end{equation}

\noindent where $G(p) = R + Mg^2(p)$, $G'(p) = \frac{\partial G(p)}{\partial p}$ and $F(p)$ is the arterial blood pressure 
distribution. The physiologically viable solution for equation \ref{eq:noise_induced} must have 
$F(p \rightarrow \infty)=0$ as boundary condition. In those conditions, equation \ref{eq:noise_induced} 
has a \textit{q}-Gaussian distribution as stationary solution \cite{PhysLettA:245:67}:

\begin{equation}
F(p) = N\left[1 - (1-q) \beta g(p)^2\right]^{\frac{1}{1-q}}
\label{eq:qgaussian}
\end{equation}

By replacing equation \ref{eq:qgaussian} in equation \ref{eq:noise_induced}, we obtain:

\begin{equation}
 gg' \{  ( \tau + M ) [ 2 + \beta \tau (q-1)g^{2}  + 2 ] - \beta \tau ( R + M g^2 ) \} = 0.
 \label{eq:cond0}
\end{equation}

If $gg' = 0$ implies $f(p) = 0$ for some $p^*$.  As $A < C$ and $B > 0$ then $p^* = -\frac{1}{B} ln( \frac{C}{A} -1)> 0$. 
The positive value of $p^*$ is associated with the continuous action of the 
sympathetic activity when not inhibited. In other words, when the blood pressure is near the central value there is an continuous 
activity, called sympathetic tone, of the nervous system to increase the blood pressure.
Otherwise, if $gg' \neq 0$, equation \ref{eq:cond0} must hold for every $p$, which implies $q =2 - \frac{1}{1+M/\tau}$ and 
$\beta = \frac{2(1+M/\tau)}{R}$.
If coupling constant $\tau$ is larger than multiplicative noise amplitude such as $M/\tau \rightarrow 0$ then 
$q \rightarrow 1$ and $F(p) $ converge to Gaussian distribution. The multiplicative noise 
becomes too small and the long tail vanish. 
On the other hand, if $M/\tau \rightarrow \infty$ $q \rightarrow 2$ and $F(p)$ converges to the Lorentz distribution \cite{umarov2008q} 
destroying homeostasis.
In spite of \textit{q}-Gaussian distribution are defined for $q<3$, to describe the non-Gaussian behavior observed in
figure \ref{fig:represent}c and keep the physiological feasibility, \textit{q} must stick in between 1 and 2. 

To compare the proposed model with the experimentally observed diastolic blood pressure data, a rescale transformation 
where performed: $p_s = k_0(p - \tilde{p})/\sigma$, where $k_0 = 4$, $p$ is the blood pressure, $\tilde{p}$ is it median 
and $\sigma$ is it standard deviation. 
Once the animals were unrestrained, each BP series could shown different central values and variability. However, when the rescale 
where performed the data collapse as figure \ref{fig:colapse} shows. In the same figure the stable solution of the model, equation \ref{eq:qgaussian},
where plotted. For control animals, figure \ref{fig:colapse_rats_ctr}, the following parameters were used: $\tau = 21/5$, $M=6/5$, $A=7/5$, $B=2/5$, and $C=9/5$. 
These values implying $q=11/9 \approx 1.22$. A value of $q=1.26 \pm 0.1$  where observed in 
heart rate variability \cite{PhysA.395.58}. Several 
other non-equilibrium systems presents $q=1.22$,for example, finacial markets \cite{ausloos2003dynamical}, hadron-hadron collisions 
\cite{barnafoldi2011tsallis}, and geological faults \cite{vallianatos2013non}. For surgically disrupted animals, presented at figure 
\ref{fig:colapse_rats_20d}, the following parameters were used: $\tau = 26/5$, $M=1/5$, $A=1$, $B=1/5$, and $C=9/5$. These values implying 
$q=1.04$, very close to the Gaussian distribution ($q=1.0$), showing that the non Gausssian fluctuations are intrinsicly associated 
with the control feedback loop at short time scale.

\begin{figure}[h]
  \centering
  \begin{minipage}{0.99\columnwidth}
   \centering
  
  \subfloat[DFA control rats]{
      \includegraphics[height=0.7\columnwidth,angle=270]{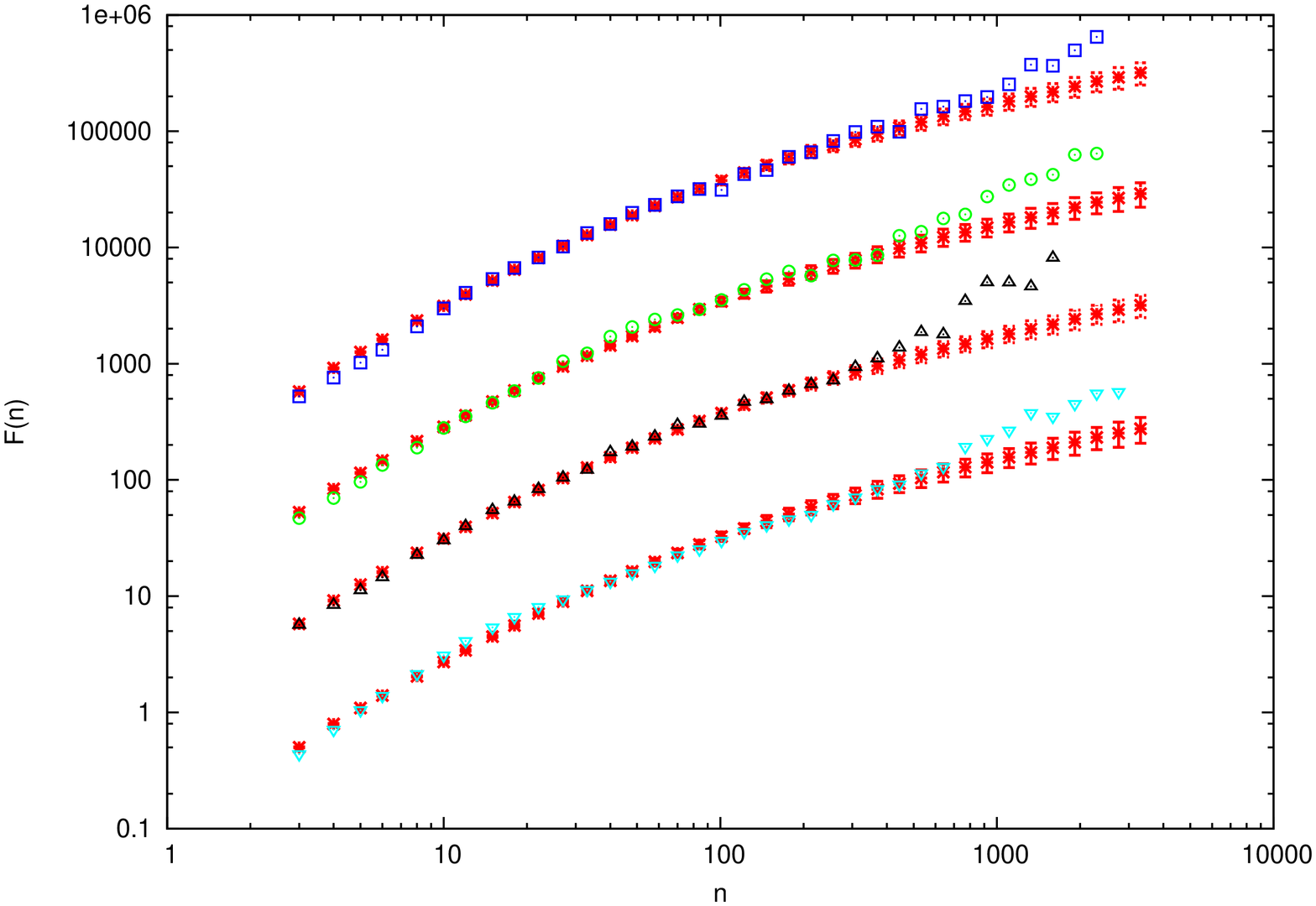}
      \label{fig:dfa_rats_ctr}
  }\\
  \subfloat[DFA denervated rats]{
      \includegraphics[height=0.7\columnwidth,angle=270]{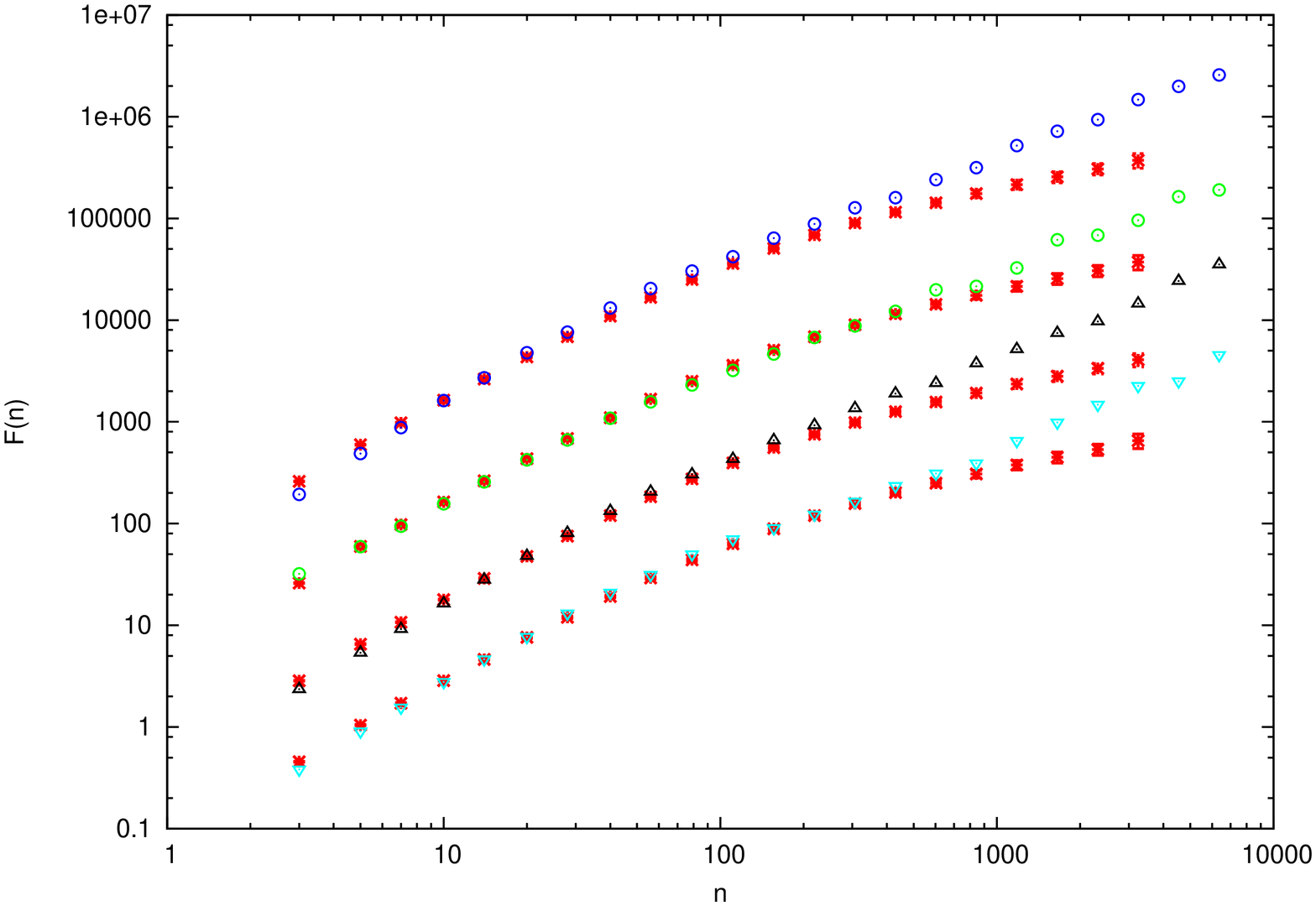}
      \label{fig:dfa_rats_20d}
  }\\
  
\caption{Comparison between DFA analysis of the experimental data and DFA of the time series 
generated by equation \ref{eq:langevin_geral}. The experimental data for four different animals of each group are plotted 
with open symbols (purple $\square$, green $\circ$, black $\vartriangle$ and blue $\triangledown$). 
The DFA from simulated series are plotted with red $\times$.  The curves were translated for better visibility, 
and values of $F(n)$ are vertically shifted. In figure \ref{fig:dfa_rats_ctr} the control group is presented while in 
\ref{fig:dfa_rats_20d} the denervated group is presented. The model capture the fluctuations behavior until the $n \approx 500$ for both 
groups. The error bars represent the standard deviation over 1000 simulations.}
\label{fig:dfa_rats}
\end{minipage}

\end{figure}

To characterize quantitatively the dynamics, a Monte Carlo simulation of the equation \ref{eq:langevin_geral} were performed for both group. 
The same parameters of the analitical curves in figure \ref{fig:colapse} were used. To understand how the model could reproduce 
the underlying dynamics a DFA of the simulated series were compared with the blood pressure real data. 
As recently discussed DFA could present a biased estimator for the Hurst exponent \cite{bryce2012revisiting}. 
Nevertheless it still holding as good methodology to work on real data \cite{shao2012comparing}, and any bias that 
could be introduced in the fluctuation function will happens on both time series, the real and 
the simulated one, being no hindrance to the analysis. 
The results of the DFA for synthetic series and real data is presented in figure \ref{fig:dfa_rats}.
The time scales of the blood pressure control feedback loop is commonly analyzed in three different ranges: 
high , low and very low frequency . The baroreflex plays a significant
role at high and low frequency control \cite{cerutti1994baroreflex}. 

After 20 days, since major damages were inflicted on the neural circuit responsible for the baroreflex, 
the control were achieved by some other redundant mechanism but with larger response delay \cite{NJP:11:103005}. 
Figure \ref{fig:dfa_rats} shows the model captures the fluctuations behavior in the baroreflex control range for both groups. 
Time scales larger than $n > 500$ are in the very low frequency range and the model developed here plays no role. 
However, except for long time scales, the fluctuation functions reveals a very similar pattern between synthetic and experimental 
time series. Based on this analysis, we conclude the model presented here captures the complex behavior of blood pressure control. 

A langevin model based on a sigmoidal restoring force with multiplicative noise for the diastolic blood pressure 
time series was discussed. The stationary solution was a q-Gaussian distribution with $q=11/9$ that 
describes remarkably well the blood pressure time series 
recorded from femoral artery for $30$ minutes in conscious unconstrained rats.
To investigate the model dynamics a synthetic time series was generated 
by a monte carlo simulation of the equivalent Langevin equation. The synthetic time series and the 
experimental data was characterized using DFA and the fluctuation functions were compared. 
The fluctuation functions reveals a very similar pattern in the synthetic and experimental time series for high and 
low frequency. 


\bibliographystyle{apsrev4-1} 
\bibliography{gralha}
\end{document}